\documentclass[10pt]{iopart}

\usepackage{graphicx}
\usepackage{dcolumn}
\usepackage{bm}
\usepackage[T1]{fontenc}
\usepackage[latin1]{inputenc}
\usepackage[english]{babel}
\usepackage{color}
\usepackage{pgf}
\usepackage{multimedia}
\usepackage{mathptmx}
\usepackage{mathrsfs}
\usepackage{amsfonts}
\usepackage{iopams}
\usepackage{selinput}
\usepackage{ulem}

\begin{document}

\title{Thermopower and thermal conductivity in the Weyl semimetal NbP}
\author{U~Stockert$^1$, R~D~dos~Reis$^{1,\footnote{\rm Present address: Brazilian Synchrotron Light Laboratory (LNLS), Brazilian Center for Research in Energy and Materials (CNPEM), 13083-970, Campinas, Sao Paulo, Brazil}}$, M~O~Ajeesh$^1$, S~J~Watzman$^2$, M~Schmidt$^1$, C~Shekhar$^1$, J~P~Heremans$^{2,3,4}$, C~Felser$^1$, M~Baenitz$^1$ and M~Nicklas$^1$}

\address{
$^1$ Max Planck Institute for Chemical Physics of Solids,  N\"{o}thnitzer Str.\ 40, 01187 Dresden, Germany \\
$^2$ Department of Mechanical and Aerospace Engineering, The Ohio State University, Columbus, OH 43210, USA \\
$^3$ Department of Physics, The Ohio State University, Columbus, OH 43210, USA \\
$^4$ Department of Materials Science and Engineering, The Ohio State University, Columbus, OH 43210, USA \\
}

\ead{nicklas@cpfs.mpg.de}

\begin{abstract}
The Weyl semimetal NbP exhibits an extremely large magnetoresistance (MR) and an ultra\textcolor{black}{-}high mobility. The large MR originates from \textcolor{black}{a} combination of the nearly perfect compensation between electron- and hole-type charge carriers and the high mobility, which is relevant to the topological band structure. In this work we report on temperature\textcolor{black}{-} and field-dependent thermopower and thermal conductivity experiments on NbP. Additionally, we carried out complementary heat capacity, magnetization, and electrical resistivity measurements. We found a giant \textcolor{black}{adiabatic} magnetothermopower with a maximum of $800~{\rm \mu V/K}$ at 50~K in a field of 9~T. Such large effects have been observed rarely in bulk materials. We suggest that the origin of this effect might be related \textcolor{black}{to} the high charge-carrier mobility. \textcolor{black}{We further observe} pronounced quantum oscillations in both thermal conductivity and thermopower. The obtained frequencies compare well with our heat capacity and magnetization data.
\end{abstract}

{\hspace{18mm}\tiny\scriptsize{\today}}

\submitto{\JPCM}

\maketitle

\ioptwocol

\section{Introduction}

The Weyl semimetals exhibit both topological surface states characterized by Fermi arcs and linear dispersive chiral bulk states\textcolor{black}{,} which lead to exotic transport properties such as the chiral anomaly \cite{Adler1969,Bell1969,Wehling} and an extremely large transversal magnetoresistance (MR) \cite{Shekhar2015}. The recent discovery of these behaviors in transition-metal monopnictides attracted \textcolor{black}{significant} attention \cite{qi2011RMP,Hasan:2010ku,Yan2012rpp,Weng2015,Huang2015,Xu2015TaAs,Lv2015TaAs,Yang2015TaAs,Arnold16}. The band structure of a Weyl semimetal possesses pairs of bands crossing at isolated points in the Brillouin zone, the so-called Weyl points. \textcolor{black}{Unlike graphene, which has two dimensional (2D) bands \cite{Neto}} , the Weyl semimetals have a three dimensional (3D) band structure \cite{Turner:2013tf,Hosur:2013eb}. This three dimensionality has important consequences for the stability of the Weyl points \cite{Wehling}.

In particular, NbP exhibits an extremely large transversal MR and an ultra high charge-carrier mobility \cite{Shekhar2015}. The Fermi surface of NbP consists of a pair of spin-orbit-split electron pockets at the Fermi energy and a similar pair of hole pockets in which one pocket nests inside the other one like a Matryoshka doll for both electron and hole pockets \cite{Klotz_2016,Sergelius}, but with persistent linear energy dispersion. NbP possesses two groups of Weyl nodes, W1 and W2 at $-57$~meV ($\approx 661$~K) and $5$~meV ($\approx 58$~K) away from the Fermi energy, respectively \cite{Klotz_2016}. \textcolor{black}{Furthermore, it was shown} that their calculated position with respect to the Fermi energy is extremely sensitive to tiny changes in the lattice parameters due to the reduced density of states \cite{Reis_2016}.

To identify candidate materials for a Weyl semimetal state, it is important to \textcolor{black}{characterize their properties fully}. The thermoelectric power $S$ is a powerful tool to probe electronic relaxation processes in metals and semiconductors since it provides complementary information to other techniques. The thermopower can be expressed as $S=S_{d} + S_{g}$, \textcolor{black}{where} $S_{d}$ is the contribution from \textcolor{black}{charge carrier} diffusion processes\textcolor{black}{,} and $S_{g}$ is the term related to phonons, which is typically small. The Mott formula provides a description of the diffusion part of the thermopower, where $S_{d}(B)/T$ ($B$ is the magnetic field and $T$ the temperature) depends on the energy derivative of the density of the states $N(E)$ and of the mobility $\mu(E)$ around the Fermi level, $S_{d}(B)/T\approx\partial N/\partial E +\partial \mu/\partial E$ \cite{Jefrey,Pal}. Whereas for a 2D \textcolor{black}{Dirac semimetal}, like graphene, $N(E)\propto~E$ \cite{Neto}, in a 3D Dirac or Weyl semimetal, $N(E)\propto~E^{2}$ is expected for the point nodes \cite{Wehling}. \textcolor{black}{First,} this would lead to a $T$-linear behavior in $S_{d}/T$, if $N(E)$ provides the dominant term. Since the Weyl node W2 \textcolor{black}{in NbP} is about 60~K above the Fermi energy\textcolor{black}{,} the low temperature behavior might be dominated by trivial bands and not by Weyl fermions. In the monopnictides, such as NbP, the mobility shows a strong dependence on temperature. This makes it experimentally almost impossible to disentangle the contributions from the density of states and from the mobility. Moreover, we have a multi-band scenario for which the electrons and hole charge carriers partially compensate each other. Even though there have been some theoretical efforts to predict the thermal properties of Weyl semimetals \cite{Wang, Sun2015,Ashby,Lundgren}, only \textcolor{black}{a} few experimental studies have been reported so far \cite{Lundgren,Ong,Lia,Sarah}.

In this work, we performed thermopower and thermal conductivity  experiments on the Weyl semimetal NbP, which were complemented by magnetization, electrical resistivity\textcolor{black}{,} and heat capacity measurements.  We found that NbP exhibits a giant \textcolor{black}{adiabatic} magnetothermopower with a maximum of $800~{\rm \mu V/K}$ at 50 K in a field of 9~T. \textcolor{black}{Pronounced quantum oscillations are visible in both, thermal conductivity and thermopower.
The corresponding frequencies from analyzing the thermopower and thermal conductivity compare well with the frequencies obtained from magnetization and heat-capacity data.}


\section{Experimental details}

High-quality single crystals of NbP were grown via chemical vapor transport reaction. More details on the sample preparation and characterization can be found in Ref.\ \cite{Shekhar2015}. For the thermopower and thermal conductivity experiments\textcolor{black}{,} we investigated a single crystalline NbP sample with dimensions of about $1.5 \times 0.6 \times 0.19~{\rm mm^3}$. The magnetic field \textcolor{black}{dependences} of the thermal conductivity and thermopower were recorded for selected constant temperatures between 2~K and 290~K using a modified sample holder for the thermal transport option of a PPMS (Quantum Design) (see inset of figure~\ref{SvsB}). The instrument applies a relaxation-time method with a low-frequency\textcolor{black}{,} square-wave heat pulse generated by a resistive heater. The temperature difference along the sample is measured by two calibrated bare-chip Cernox sensors (Lakeshore). \textcolor{black}{One end of the sample was attached to the heat sink using silver paste. Thermal contacts to the heater and the thermometers were accomplished via Au wires. This widely used way of sample mounting with the heat current $q$ applied along the long direction $x$ of the sample yields the thermal conductivity $\kappa_{\parallel x}$ and adiabatic thermoelectric power $S_{\parallel x}$. In general, a non-ideal contact geometry, i.e., a misalignment of the voltage leads, can give rise to a Nernst contribution in the thermopower data. Because the Nernst effect is generally an odd function of magnetic field, this contribution can be eliminated by symmetrizing the data taken in positive and negative field, which we have done. In most materials, the difference between the isothermally and the adiabatically determined thermoelectric power is extremely small. However, in NbP the temperature gradient $\nabla T$ might not necessarily be parallel to $x$ in magnetic fields for adiabatically measured data due to the Righi-Leduc effect. In that case, the magnetic-field dependence of the Nernst coefficient also folds into the measurement of the adiabatic thermopower \cite{Sarah}. This effect cannot be eliminated by data post-processing.}

\textcolor{black}{Thermopower and thermal conductivity} data were obtained at constant field and temperature as an average over 10 individual heat pulses. In all measurements the heat current was applied along the $[010]$ and the magnetic field along the $[001]$ axes of the tetragonal crystal structure of NbP. We note that\textcolor{black}{,} due to the strong temperature dependence of $\kappa$, a measurement of the temperature dependence of the thermal conductivity and thermopower while sweeping the base temperature was not successful in the whole temperature range. However, for low ($T < 10$~K) and high ($T > 100$~K) temperatures, the results obtained during temperature sweeps agree with those obtained during field-dependent measurements at constant temperature. The magnetization and heat capacity experiments were performed at several temperatures in magnetic fields up to $B=9$~T applied along the $[001]$ axis in a PPMS.


\section{Results and discussion}

\begin{figure}[tb]
\begin{center}
\includegraphics[width=0.4\textwidth]{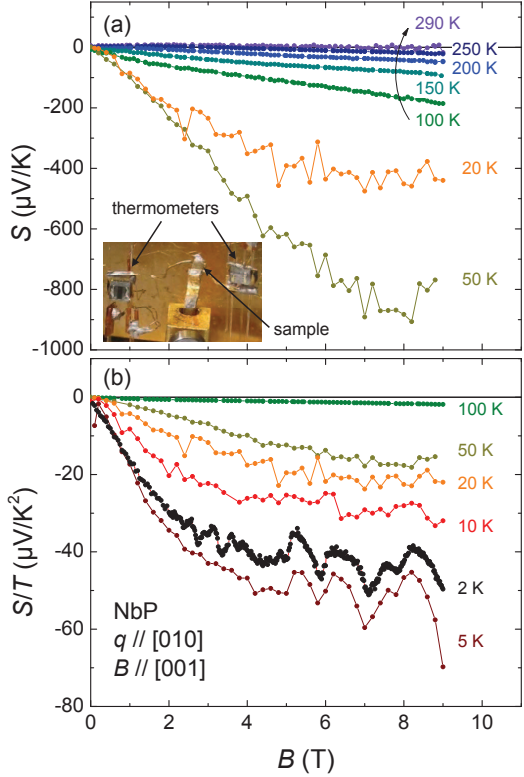}
\end{center}\caption{(Color online) Field dependence of the \textcolor{black}{adiabatic} thermopower of NbP at different temperatures, plotted as (a) $S(B)$ in the temperature range ($T \geq 20$~K) and (b) as $S(B)/T$ for $T \leq 100$~K. Towards low temperatures\textcolor{black}{,} the emergence of quantum oscillations is clearly visible. The inset in (a) shows a photograph of the modified thermal transport option of the PPMS for thermal conductivity and thermopower experiments.}
\label{SvsB}
\end{figure}

The extremely low effective masses and the high mobilities of the charge carriers make NbP an ideal platform to explore the thermoelectric response of a Weyl semimetal. We start our discussion with the magnetic-field dependence of the \textcolor{black}{adiabatic} thermopower $S(B)$ taken at different temperatures. Figures \ref{SvsB}(a) and \ref{SvsB}(b) present data for $T \geq 20$~K as $S(B)$ and for $T \leq 100$~K as $S(B)/T$, respectively.
The absolute values of the thermopower $\left|S\right|$ are strongly enhanced \textcolor{black}{with} increasing magnetic field. This behavior is observed for all temperatures below 200~K. In the high temperature range above 100~K, the thermopower depends almost linearly on the magnetic field, while it shows a tendency to saturation in large fields at lower temperatures. \textcolor{black}{Below 20~K, we start to} observe  signatures of quantum oscillations in the data, which are directly related to the topology of the Fermi surface of NbP \cite{Klotz_2016}. Due to the small size of the oscillatory component of the signal and the small number of data points, they are difficult to identify. \textcolor{black}{However, at 2~K,} the resolution is sufficient and due to smaller field steps in the measurements, the oscillatory part of the data can be \textcolor{black}{seen clearly}. A detailed analysis of the quantum oscillations will follow below. At temperatures $T\leq 100$~K, a more appropriate presentation of the thermopower effect is a plot of $S(B)/T$ (see figure~\ref{SvsB}(b)), since $S(T)/T$ is supposed to be \textcolor{black}{temperature independent} in a conventional \textcolor{black}{Fermi liquid} at low $T$. \textcolor{black}{We observe a consistent} behavior in small magnetic fields below 1~T. At higher fields\textcolor{black}{,} $\left|S(T)\right|/T|_{B={\rm const.}}$ first increases upon decreasing temperature to 5~K before it decreases again towards $2$~K.

At first glance, the thermopower of NbP changes smoothly with magnetic field and temperature and does not show any unusual or abrupt behavior. However, the magnitude of the change induced by a magnetic field is huge compared to most other materials. The magnetothermopower defined as $\left|S(B)-S(0)\right|$ reaches values as large as $800~\mu$V/K in a field of 9~T at 50 K. This value \textcolor{black}{is more} than one order of magnitude larger than that observed in semimetals such as graphite \cite{Woollam}. \textcolor{black}{Larger values up to $1000~\mu$V/K in 1.85 T have been reported for doped InSb \cite{Here01a, Here01b} and attributed to the presence of electron- and hole-type charge carriers with different mobilities and to a geometrical magnetothermopower.} A remarkably large thermopower effect associated with the high mobility of the charge carriers was also reported in Bi ($4000~\mu$V/K at 5~T and 7~K) \cite{Mangues}. Other materials with \textcolor{black}{``giant''} magnetothermopower include colossal magnetoresistance manganites ($20~\mu$V/K at 150~K and 8~T) \cite{Jaime}, the doped semiconductor AgTe (470~$\mu$V/K at 110~K and 7~T) \cite{Sun03a}, charge ordered Nd$_{0.75}$Na$_{0.25}$MnO$_3$ (50~$\mu$V/K at 80~K and 3~T) \cite{Repa13a} and PtSn$_4$ (45~$\mu$V/K at 15~K and 14~T) \cite{Mun12a}. \textcolor{black}{Therefore, the observation of this large magnetothermopower of more than 800~$\mu$V/K in NbP raises the question of its origin.}

\begin{figure} [t!b]
\begin{center}
\includegraphics[width=0.45\textwidth]{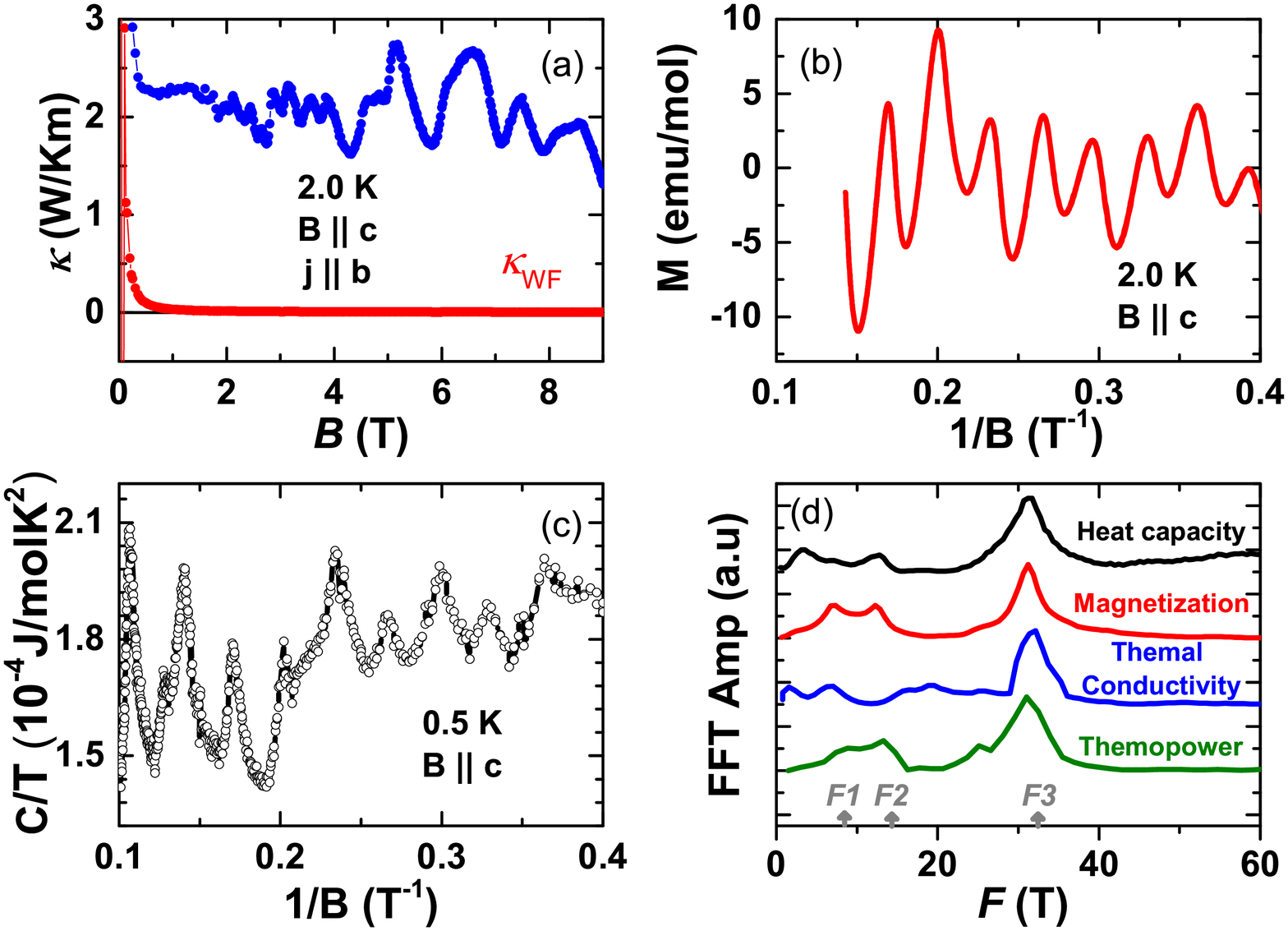}
\caption{(a) Field dependence of the thermal conductivity $\kappa$ of NbP at 2~K compared to the contribution $\kappa_{\mathrm{WF}}$ estimated from the simultaneously recorded electrical resistivity using the Wiedemann-Franz law. (b) Magnetization $M$ at $T=2$~K and (c) the heat capacity as $C/T$ at $T=0.5$~K as a function of the inverse of the magnetic field $1/B$. (d) Results of a fast Fourier transformation of heat capacity, magnetization, thermal conductivity\textcolor{black}{,} and thermopower data. The arrows labeled as $F1$ (8.7~T), $F2$ (14.2~T)\textcolor{black}{,} and $F3$ (32.1~T) represent the experimental frequencies for electrons orbits for $B\parallel c$, previously reported in the literature \cite{Klotz_2016}.}
\label{QO_NbP}
  \end{center}
\end{figure}

Now, we turn to the \textcolor{black}{adiabatic} thermal conductivity $\kappa$ of NbP. Figure \ref{QO_NbP}(a) shows the magnetic-field dependence of $\kappa$ at 2~K. Two different regimes can be distinguished: for small fields, $B < 0.5$~T, a strong drop of the thermal conductivity is observed. In fact, we were not able to accurately determine the zero-field thermal conductivity at 2~K due to the extremely small temperature gradient that could be realized without a significant heating of the sample. $\kappa$ could \textcolor{black}{be} estimated to be within 45 and 70~W/Km. \textcolor{black}{At higher fields, we observe strong quantum oscillations in $\kappa(B)$ sitting on an almost-constant background. Note that strong quantum oscillations also are present in the electrical resistivity \cite{Shekhar2015}. Quantum oscillations in the thermal conductivity are of purely electronic origin \cite{Bhagat}. Therefore, they are expected only in the electronic part of the thermal conductivity. In semimetals, the electronic contribution to the thermal conductivity consists of three parts: $\kappa_{el}=\kappa_{e}+ \kappa_{h}+ \kappa_{amb}$, where the first and second terms are the ordinary thermal conductivity due to electrons and holes, respectively.  The third term represents the ambipolar diffusion term, which is a contribution to the electronic thermal conductivity made by the thermal energy released when an electron and hole annihilate one another after traversing a temperature gradient \cite{Ziman}. $\kappa_{e}$ and $\kappa_{h}$
are connected to the electrical conductivity by the Wiedemann-Franz law $\kappa_{\mathrm{WF}}/\sigma=LT$, where $\sigma$ is the electrical conductivity and $L=\frac{\pi^2}{3}\left(\frac{k_B}{e}\right)^2$ the Lorenz number. We can estimate this contribution from the electrical resistivity of the sample, which was determined simultaneously with the thermal conductivity using the same experimental leads and contacts. We find that $\kappa_{\mathrm{WF}}(B)$ in NbP is at least two orders of magnitude smaller than the experimentally measured $\kappa(B)$, except for very low fields (see figure~\ref{QO_NbP}). Therefore, this contribution cannot account for the strong quantum oscillations in the thermal conductivity. Thus, we conclude that the electronic contribution to the thermal conductivity is dominated by the ambipolar term \cite{Ziman}. In the following, we will compare the results on the quantum oscillations in the thermal conductivity with magnetization and heat capacity data (see figures \ref{QO_NbP}(b) and \ref{QO_NbP}(c)).}

In order to obtain the frequencies of the quantum oscillations, we subtracted an appropriate polynomial background from the corresponding data. The frequencies were then determined by a fast Fourier transform (FFT) on the oscillatory part of the signal. The FFT amplitude of the quantum oscillations in our heat capacity, magnetization, thermal conductivity\textcolor{black}{,} and thermopower data are shown in figure~\ref{QO_NbP}(d). Additionally, we have indicated the three frequencies related to the electron pockets, labeled as $F1$, $F2$\textcolor{black}{,} and $F3$, which were obtained in the same frequency interval by high\textcolor{black}{-}precision torque experiments \cite{Klotz_2016}. $F3$ is well\textcolor{black}{-}resolved in the data obtained by the different physical probes, \textcolor{black}{while the smaller frequencies $F1$ and $F2$ cannot be \textcolor{black}{detected unambiguously} by all techniques due to the different experimental resolution of each technique.} This also precludes resolving all frequencies in the accessible range. The slightly smaller frequencies obtained in our experiments compared with the values from literature \cite{Klotz_2016} are most likely associated with a small misalignment between the direction of the magnetic field and the crystallographic $c$-axis of the sample. This is consistent with the reported angular dependence of the observed frequencies \cite{Klotz_2016}.

\begin{figure}[tb]
\begin{center}
\includegraphics[width=0.45\textwidth]{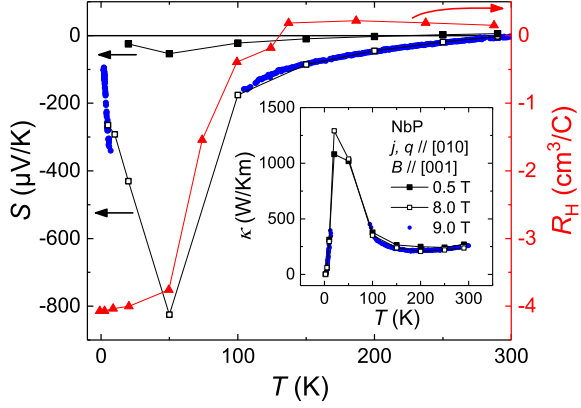}
\end{center}
\caption{(Color online) \textcolor{black}{Adiabatic} thermopower $S(T)$ in magnetic fields of 0.5~T and 8.0~T (left axis) estimated from the corresponding field-dependencies depicted in figure~\ref{SvsB} by averaging over a field interval of $\pm 0.3$~T. Hall coefficient $R_H$ (right axis) taken from Ref.~\cite{Shekhar2015}. The inset displays the thermal conductivity $\kappa(T)$. Additionally, thermopower and thermal conductivity data recorded during a temperature sweep in a field of 9~T are shown. Here, it was not possible to obtain data at intermediate temperatures due to the strong temperature dependence of $S$ and $\kappa$ as explained in the experimental part.
}
\label{SvsT}
\end{figure}

For a \textcolor{black}{simple,} single band \textcolor{black}{model,} the type of charge carriers is \textcolor{black}{related directly} with the sign of \textcolor{black}{the} thermoelectric effect. In the case of NbP, the material hosts electron- and hole-type carriers. Furthermore, the concentration of both exhibits a strong temperature dependence (see figure~\ref{SvsT}) \cite{Shekhar2015}. In particular, the type of the dominant charge carriers changes at about 125~K\textcolor{black}{,} as evidenced by the Hall coefficient $R_H$ (see figure~\ref{SvsT}) \cite{Shekhar2015}. In NbP\textcolor{black}{,} the negative $R_H$ below 125~K indicates predominant electron-type charge carriers, while above 125~K\textcolor{black}{,} $R_H$ changes sign\textcolor{black}{,} indicating predominant hole-type charge carriers. The temperature dependence of $S(T)$ for $B=0.5$, 8\textcolor{black}{,} and $9$~T is depicted in figure~\ref{SvsT}. For $B=0.5$~T\textcolor{black}{,} the thermopower is relatively small in the entire temperature range\textcolor{black}{;} only at $T=50$~K \textcolor{black}{does a shallow dip appear}. In contrast to that, at $B=8.0$~T, $S(T)$ displays a surprisingly deep minima at the same temperature. Upon decreasing the temperature\textcolor{black}{,} $\left|S(T)\right|$ increases up to as much as $800~\mu$V/K at 50~K, followed by a steep decrease down to less than $100~\mu$V/K at 2~K. We point out that the pronounced increase in the thermopower starts below $100$~K, in the region where the electrons become the dominant charge carriers in NbP. \textcolor{black}{Below $T\leq 50$~K, the charge-carrier concentration does not change anymore and the thermopower decreases as in a normal metal.}

\textcolor{black}{In the following,}  we come back to the huge \textcolor{black}{adiabatic} magnetothermopower observed for NbP at intermediate temperatures. \textcolor{black}{We first consider phonon drag as a possible source. In general, at low temperatures the absolute phonon-drag thermopower $\left|S_\mathrm{d}(T)\right|$ increases with increasing temperature as the phonon density increases. Upon further increasing phonon-phonon interactions at higher temperatures, $\left|S_\mathrm{d}(T)\right|$ starts to decrease again, giving rise to a maximum at intermediate temperatures. This is what we observe for the thermopower recorded on NbP in magnetic fields, as shown in figure \ref{SvsT}. However, for semimetals, such as NbP, only phonons with long wave lengths are involved in the drag processes due to the small size of the Fermi surface. \textcolor{black}{In} contrast, the entire phonon spectrum contributes to phonon-phonon scattering and, correspondingly,  to the thermal conductivity \cite{Herring}. Therefore, in semimetals, the maximum in $\left|S_\mathrm{d}(T)\right|$ is expected at much lower temperatures than the one in $\kappa(T)$, unlike it is the case in NbP. }

\textcolor{black}{The maximum phonon energy relevant for drag processes can be estimated. Only phonons with wave numbers comparable to or smaller than the Fermi-wave number are involved in the phonon-drag processes. The Fermi-wave number of NbP can be obtained from the frequencies of the quantum oscillations. The largest one $F3$ \cite{Klotz_2016}, which is also resolved in our data (see figure \ref{QO_NbP}), corresponds to a wave number of $3 \times 10^8$~m$^{-1}$. This is only about 3.2~\% of the Brillouin zone size perpendicular to $c$, which is $9.4 \times 10^9$~m$^{-1}$, calculated from the lattice parameter $a=3.33$~\AA\ \cite{Shekhar2015}. Since phonons at the Brillouin zone edge have an energy equivalent to the Debye temperature $\Theta_\mathrm{D}$ of about 455~K \cite{baenitz}, only those phonons with energies up to about 15~K may contribute to drag processes.  Therefore, a maximum in $\left|S_\mathrm{d}(T)\right|$ caused by phonon drag is expected only around or below that temperature, in contrast to our observation of a maximum in $\left|S(T)\right|$ at about 50~K. Thus, a phonon drag seems to be unlikely as the origin of the observed maximum. However, an exact theoretical treatment of the phonon-drag effect in the thermopower is complicated and requires, in particular, knowledge on the transition probabilities for electron-phonon scattering processes.}

\textcolor{black}{A large Nernst thermopower, on the order of magnitude of the magnetothermopower, has been reported in NbP, with no measurable isothermal magnetothermopower \cite{Sarah}. Based on this result, a possible explanation for the observation of the strong magnetic-field dependence in $\left|S_\mathrm{d}(T)\right|$  is related to the adiabatic magnetothermopower measurement technique used here.  The adiabatic magnetothermopower is subject to not only the longitudinal thermopower, but also to a contribution from the Nernst effect and a contribution from the Righi-Leduc effect, which induces a temperature gradient in the direction mutually orthogonal to both the applied temperature gradient and magnetic field \cite{Harman}. Conversely, in an isothermal magnetothermopower measurement, the signal is subject only to the longitudinal thermopower. }


\section{Conclusions}

In summary, we investigated the \textcolor{black}{adiabatic} thermopower and thermal conductivity of the noncentrosymmetric Weyl semimetal NbP. \textcolor{black}{We find evidence for a large, ambipolar contribution to the electronic thermal conductivity at low temperatures, indicating that the energy carried in movement of both electrons and holes in a charge-neutral environment dominates the thermal conductivity at these temperatures.} The frequencies of the observed quantum oscillations in thermal conductivity, thermopower, specific heat, and magnetization are in good agreement. The \textcolor{black}{adiabatic} magnetothermopower peaks at 50~K and exhibits a rather large value of $\left|S(T)\right|\approx 800~\mu$V/K. \textcolor{black}{At the same temperature the thermal conductivity also displays a peak.} Our results demonstrate that thermopower and thermal conductivity are important tools to study Weyl semimetals and stimulate further studies.


\section*{Acknowledgements}

We thank E.\ Hassinger, T.\ Oka and B.\ Yan for fruitful discussions and V.\ S\"{u}{\ss} for preparing the samples. RDdR acknowledges financial support from the Brazilian agency CNPq (Brazil). \textcolor{black}{SJW is supported by the U. S. National Science Foundation Graduate Research Fellowship Program under grant number DGE-0822215 and JPH is supported by the Center for Emergent Materials, an NSF MRSEC, under grant number DMR-142045.}


\begin{thebibliography}{99}
\bibitem{baenitz} $\Theta_\mathrm{D}\approx 455$~K has been estimated from our specific heat data.
\end{thebibliography}
\end{document}